# Modern Data Formats for Big Bioinformatics Data Analytics

Shahzad Ahmed
Department of Computer Science
COMSATS Institute of Information Technology
Sahiwal, Pakistan

Javed Ferzund
Department of Computer Science
COMSATS Institute of Information Technology
Sahiwal, Pakistan

Abbas Rehman
Department of Computer Science
COMSATS Institute of Information Technology
Sahiwal, Pakistan

M. Usman Ali
Department of Computer Science
COMSATS Institute of Information Technology
Sahiwal, Pakistan

Muhammad Atif Sarwar
Department of Computer Science
COMSATS Institute of Information Technology
Sahiwal, Pakistan

Atif Mehmood
Riphah Institute of Computing and Applied Sciences (RICAS)
Riphah International University
Lahore, Pakistan

***Abstract*—Next Generation Sequencing (NGS) technology has resulted in massive amounts of proteomics and genomics data. This data is of no use if it is not properly analyzed. ETL (Extraction, Transformation, Loading) is an important step in designing data analytics applications. ETL requires proper understanding of features of data. Data format plays a key role in understanding of data, representation of data, space required to store data, data I/O during processing of data, intermediate results of processing, in-memory analysis of data and overall time required to process data. Different data mining and machine learning algorithms require input data in specific types and formats. This paper explores the data formats used by different tools and algorithms and also presents modern data formats that are used on Big Data Platform. It will help researchers and developers in choosing appropriate data format to be used for a particular tool or algorithm.**

## I. Introduction

In upcoming era, a lot of data will be generated in various fields such as Computer Science, Electrical, Mechanical, Management Science, Mathematics and Bioinformatics. Today, data in these fields is being generated very quickly due to advancement in technology and development of new techniques and tools. Speed of data generation increases in seconds day by day. There is a big challenge to handle and manage such large volume of data. To overcome this problem many tools are designed that handle, manage and store the data in any field.

The term Big Data is well-defined as the data becomes too large that is difficult to handle, process, store and manage. A lot of data is generated from Twitter, Skype, Google, Facebook, Walmart, IOT applications, Yahoo and Youtube. Data has been produced in all sectors like education, professional services, banking and finance. Such large amount of data requires better storage and analysis. Big Data consists of five characteristics like Variety, Velocity, Veracity, Volume and Potential Value. The last characteristic is very vital. Real Time analysis plays an imperative role in Big Data. ML (Machine Learning), Data Mining and Cloud computing are important aspects of Big Data applications. Major challenges and risks of Big Data are cost, flexibility, accuracy, scalability and performance.

With the passage of time, data has rapidly increased in the Bioinformatics field such as public health care data, imaging data, clinical data, sequencing data, genome data and protein data. Large amount of data is also generated from relationships such as gene-disease, protein-protein and DNA-protein etc. In clinical domain, data of many patients is collected for intelligent decision making. This data needs to be stored and analyzed in an efficient way that would be helpful for making best decisions. In imaging domain, huge amount of data about medical images is collected and shared in an effective way. Imaging informatics is beneficial for patient's recovery. Public health care domain consists of prediction and finding of infectious diseases using Big Data tools and technologies [1]. All the data in these domains need better storage facility.

A Format (Model) is a predetermined layout for data. A computer program takes data as input in a certain format, processes it, and gives data after processing that is called information as output in the same or another format.

Machine Learning plays an important role in Bioinformatics field. ML (Machine Learning) Algorithms and Techniques are used for Classification and Clustering of proteins data, sequencing data, genomics data etc. A lot of ML Algorithms such as kNN (k Nearest Neighbor), SVM (Support Vector Machine), Logistic Regression, Naïve Bayes, k-means, k-median, GLM (Generalized Linear Model), Decision Tree and Random Forest are available that perform Classification and Clustering tasks for Bioinformatics datasets.

It stores data in the form of key value. There are two stages map and reduce stage.in the map phase, input data stored in HDFS (Hadoop Distributed File System). Mapper processes that data and creates multiple chunks. Then, Reducer processes that input data coming from mapper and provides a set of outputs. Output files are stores in HDFS. Many Machine Learning Classification and Clustering Algorithms are implemented in Hadoop MR. MR used mahout library for analysis of these Algorithms.

Many tools are available for large data storage, management and analysis. Most important tool is Apache Hadoop, a shared nothing open source architecture that distributes and stores huge datasets across multiple clusters. Hadoop provides features of fault tolerance, data locality, security and reliability. Hadoop components (modules) are HDFS (Hadoop Distributed File System) and MR (MapReduce). HDFS is file storage system for large datasets in Hadoop. It distributes jobs in name node and data node in reliable manner. MR is a framework that is developed for parallel execution of tasks for large datasets by using job tracker and task tracker. Many tools like HBase, Hive, Pig and Apache Spark are built on top of Hadoop. HBase is based on google big table for random read/write access to large data in NoSQL data store. It implements partition tolerance and consistency in the CAP theorem. It replaces RDBMS (Relational Database Management System) and provides the facility of automatic sharding. Hive is a system that translates MR (MapReduce) tasks. It provides the capability of large data storage in embedded, local and remote mode. Pig is used for scripts. It is executed in local and MapReduce mode [2]. It consists of many built-in-functions and operators such as arithmetic, Boolean and comparison, ORDER BY and GROUP BY for large datasets. Apache Spark is especially designed for large data analysis. It takes the benefit of our built-in libraries such as Mlib, GraphX, Spark Streaming and Spark SQL. It consists of transformations, caching and actions. It is faster than MapReduce.

In MapReduce, programming is very difficult for processing and analysis. In many circumstances, batch processing does not fit results less Performance. Spark reduces these limitations due to in-memory computations. Scala language performs superlative role in the analysis on Spark Framework. Spark also provides shell facility in java, python and scala language. There are three ways for creating RDD (Resilient Distributed Dataset) such as parallelizing, referencing and transformation from existing RDD. DAG (Direct Acyclic Graph) is very important in Spark. A lot of transformation methods are available in Spark such as map (), filter (), flatMap () and reduceByKey (). Many methods are exist in Spark for actions like count (), collect () and foreach (). Many RDD persistent storage level are available in Spark that is used according to need for analysis. Broadcast and accumulator variables are very significant for different operations.

Many data formats (Models) are available for any type of data storage. Some data formats are Comma Separated Values (CSV), Tab Separated Value (TSV), Text Format, XML, JavaScript Object Notation (JSON), Attribute-Relation File Format (ARFF), Sequence File Format and Zip file format. Specific data Format is used for specific tool for large Bioinformatics data storage. Some modern data formats (Models) also exist for large data storage that are supported by Hadoop and Apache Spark framework. These Formats are key value, Compressed Row Storage (CRS), Sparse Vector, Avro, Parquet and ORC (Optimized Row Compressed). Some Formats are used for the storage of Bioinformatics data like BAM (Binary Alignment Map), Fastq Format, FASTA Format and VCF (Variant Call Format). Some specific Formats are used for graph processing in Hadoop such as Vertex Input Format, Edge Input Format, Vertex Output Format and Edge Output Format. Many ML Algorithms and Techniques are implemented in Hadoop MapReduce and Apache Spark. These ML Algorithms are implemented in Hadoop for Bioinformatics datasets. Text Input Format is default Format for Hadoop Platform. Specific Format is used for Bioinformatics data in Hadoop MapReduce and Spark framework. Data Conversion tools are also available that convert data from one format to another specific format. These tools are Data Converters, FME data conversion and integration tool, Altova MapForce, TCS, Sqoop, Avro2parquet, Apache Drill and Format Converter.

The objectives of this study are:

- To explore the Modern Data Models (Formats) for Large Bioinformatics data storage
- To analyze the appropriate Data Models for implementation of Machine Learning Algorithms in Hadoop and Spark Framework for large Bioinformatics Data
- To present the future research directions for using Modern Data Models in the implementation of Machine Learning techniques on Hadoop and Spark Framework
- Evaluation of existing Data Models (Formats)

The rest of the paper is structured as follows: Section II enlightens the related work in this field. Section III presents Data Models (Formats). Section IV signifies Modern Data Models for the implementation of Machine Learning Algorithms in Hadoop and Spark for large Bioinformatics datasets. Section V describes tools for migration and conversion of Data Models. Section VI presents discussion about Data Models and illustrates their performance comparison.

## II. Related Work

### A. *Data Formats for Implementation of Machine Learning Techniques and Algorithms in Hadoop MapReduce*

Ali et al. [3] provide brief description of Machine Learning clustering and classification Algorithms and techniques used in big data tools such as Hadoop and Spark for large scale dataset of Bioinformatics. They also describe the performance comparison of different Machine Learning Techniques and Algorithms in the perspective of Hadoop and Spark. Abdulla et al. [4] have proposed C4.5 Decision Tree Algorithm that is implemented in MapReduce Framework using text input storage Format for measuring customer behaviour and visualization is performed using CSV (Comma Separated Value) Format by the usage of D3.js (JavaScript Library) that provide superlative results for decision making analysis. Esteves et al. [5] have developed Mahout (open source) Framework that is based on Hadoop MapReduce platform for k means, fuzzy k means and meanshift clustering for large datasets. In his work, vectors are converted to SequenceFile Format that is used as an input to the MapReduce Framework. The output files are also in the SequenceFile Format. Mahout scaling tests are performed on Amazon EC2 and best performance was produced. Stupar et al. [6] have proposed RankReduce approach that is based on LSH (Locality Sensitive Hashing) Algorithm for the implementation of multiple kNN queries on the top of MapReduce Framework. LSH will help to process the kNN queries for large datasets that provide better search accuracy results. In his work, Selected Buckets from the LSH Algorithms are used as an input to the MapReduce Framework with a lot of input splits using Input Format class and reads feature vectors are stored in BinaryFormat. Storage datasets are in BinaryFormat (Avro) for implementation of MapReduce Framework. Dai et al. [7] have developed Naïve Bayes Classification Algorithm that is combined with rough set theory for text Classification using large datasets that are implemented in MapReduce Framework. This new designed Algorithm was implemented on the cloud platform for the text Classification to achieve best accuracy and recall rate. The input is in text input Format that is given to Naïve Bayes Technique with the help of rough set. Text Input Format is used for input of preprocessing in MapReduce Framework. ZHANG et al. [8] have proposed MapReduce programming model for RBM (Restricted Boltzmann Machines) and DBN (Deep Belief Nets) in Deep Learning Technique to achieve good performance, accuracy and scalability for large datasets. RBM and DBN contains layers in which MapReduce tasks are implemented with the usage of Key Value Format within each layer. Results show that less training time is required with distributed RBM and DBN than simple RBM and DBN respectively. Ericson et al. [9] have developed Granules and Hadoop comparison platform in which Classification Algorithms (Complementary Bayes and Naive Bayes) and Clustering Algorithms (Latent Dirichlet Allocation, k means, Dirichlet and Fuzzy k means) supported by Mahout that is implemented for large datasets. Mahout used SparseVector Format for storing data internally. Results show that Granules system perform better than Hadoop platform for same Mahout Benchmark. Al-Madi et al. [10] have proposed MR-CGSO (glowworm swarm optimization Clustering) that is implemented in Hadoop MapReduce Framework using k-Median Clustering for large datasets that produce better Scalability, Speed and Accuracy. Key Value Format is used to store datasets in MapReduce Framework. Sarwar et al. [11] have proposed review study about Bioinformatics Tools. They demonstrate the implementations of Tools for Alignment Viewers, Database Search and Genomic Analysis on Hadoop and Apache Spark Framework using Scala language.

### B. *Data Formats for Implementation of Machine Learning Techniques and Algorithms in Apache Spark*

Yaqi et al. [12] have proposed Apache Spark Framework and implemented the Machine Learning Algorithm kNN using that Framework. Aliyun E-MapReduce cloud Framework is used for insulator leakage current data. Spark-kNN is faster than MapReduce. Monitoring data managed with the help of RDD (Transformation and Actions). In the Spark-kNN Algorithm, input training data is used in CSV Format and output is also in CSV Format. Results show that Spark-kNN perform better for large datasets and reduce the execution time. Armbrust et al. [13] have proposed Apache Spark module Spark SQL which include Catalyst that provide Machine Learning features. Spark SQL has strong integration with MLIB such as to train logistic regression with the help of HashingTF (frequency featurizer) to get a feature vector. It consists of relational processing. Spark SQL data storage is in columnar Format (Parquet). Bifet et al. [14] have developed StreamDM on the top of Spark streaming Library that is used in Data Mining and Machine Learning for real time analysis. A lot of stream data such as sensor and credit card transactions stored and analyzed in an efficient manner. For this purpose, many Machine Learning Algorithms and Techniques such as Naïve Bayes, Decision Trees and Bagging classifier are implemented in Spark using StreamDM Library. In his work, sparse instances stored in LibSVM Format and dense instances are stored in CSV text Format. Meng et al. [15] have developed Spark Machine Learning Library (MLIB) for many Classification and Clustering Algorithms. MLIB support LIBSVM data storage Format for the implementation of Machine Learning Techniques and Algorithms such as Decision Tree, k-means, linear models and Naïve Bayes. In [16], a model has been developed that is known as GLM (Generalized Linear Model) in SparkR on the Apache Spark Framework for airline dataset. R language and Learning Library (MLIB) are used for Machine Learning. Preprocessing of dataset are stored in CSV Format by using "spark csv" package for evaluation of large scale GLM.

### C. *Data Formats for Implementation of Machine Learning Techniques and Algorithms in Hadoop for Bioinformatics*

Ravi. R et al. [17] have proposed a system in which Machine Learning Algorithm "Scalable Random Forest" is used to select and classify the appropriate features for large datasets. This Algorithm was implemented in Hadoop MapReduce Framework by using many-many data linkage. In his work, tabular data is stored in CSV (Comma Separated Value) file Format (consisting of many records) in the plain-text form (sequence of characters). File is accessed by using blobstore (Google App Engine objects). By using many-many model, best Performance was achieved with the usage of sample dataset having 29567 rows. Karim et al. [18] have developed a method for Association Rule Mining BMR (Bio-MapReduce Framework) for large Microarray datasets. For the implementation of BMR Algorithm, writer use key value Format for input and output. Results show that BMR method performs better than traditional Algorithms. Rehman et al. [19] have explained importance of Scala language for Bioinformatics Tools/ Algorithms. They demonstrates the supported languages for Motif Finding Tools, Multiple Sequence Alignment Tools and Pairwise Alignment tools.

## III. Data Models (Formats)

A lot of Data Formats and Data Models are available for storage of big data in efficient way.

A Format for representing a data set should be:

- Compact enough to save storage for large data sets.
- Readable by various tools or softwares.

- The Format represents all categorical and numerical features so the Data Format should be rich enough.
- Be local so that parts of the data can be independently transferred.
- Data Format should not be changed capricious by tools or softwares.

There are several Formats used to represent the data sets such as CSV, TSV, TEXT, JSON, ARFF, Sequence file Format etc.

### A. *Comma Separated Values (CSV):*

In relatively short amount of time data can be swapped and transformed using several spreadsheet tools, in Comma Separated values Format. Each and every record is in one line commas separate fields from each other and prominent and irregular whitespaces are ignored. It is commonly used Format in Hadoop platform. CSV format is described in Figure. 1.

```
Title, Author, ISBN14, Page
1990, Shahzad, 978-0563457465, 67
Brave new world, Atif, 978-0564657465, 676
Eats shoots and leaves, Abbas, 97-0563657465, 66
```

Fig. 1. Comma separated Values

### B. *Tab Separated Value (TSV):*

Tab separated value is common method to exchange data among spreadsheets, databases and word processor with Mail-merge functions. Each and every record is illustrated as a single line. Every single field value is represented as a text. Fields in a record or file are delimited from each other by a tab character [20]. TSV format is described in Fig. 2.

| Sample Peak | Sample File Name | Size | Height | Area | Data | Point |
|---|---|---|---|---|---|---|
| “A, 2” | point.fsa | 13.43 | 30 | 340 | 1340 | |
| “A, 20” | point.fsa | 15.56 | 55 | 480 | 1373 | |
| “A, 21” | point.fsa | 19.56 | 35 | 3067 | 1391 | |
| “A, 22” | point.fsa | 12.60 | 40 | 120 | 1402 | |

Fig. 2. Tab separated values

### C. *Text Format:*

In plain text data is represented only in readable characters while graphical representation and other objects such as images can’t be represented in text Format. There are limited number of characters available to control common arrangement of text such as tabular characters and line breaks. Text format is described in Fig. 3.

```
package javaapplication4;
import java.util.Scanner;

public class Usman {


    public static void main(String[] args) {

       int a;
      float b;
      String s;

      Scanner i = new Scanner(System.in);
            System.out.println("Enter a string");
      s = i.nextLine();
            System.out.println("Enter an integer");
      a = i.nextInt();
            System.out.println("Enter a float");
      b = i.nextFloat();

      System.out.println("You entered string "+s);
      System.out.println("You entered integer "+a);
      System.out.println("You entered float "+b);


    }
}
```

Fig. 3. Text format

### D. *XML:*

Extensible Markup Language (XML) used to prepare Formats of data which can be easily shared over World Wide Web or anywhere using standard ASCII text. Log files mostly available in the form of XML file Format [21]. XML format is described in Fig. 4.

```
<?xml version=”1.0” encoding=”utf—8” standalone=” yes”?>
<Table >
<product >
<product_id>2</product_id>
<product_name>product 2</product_name>
<product_price>3000</product_price>
</product>
```

Fig. 4. XML format

### E. *JSON:*

JavaScript Object Notation (JSON) is used as trivial data swapping Format among different applications and tools. It is relatively easy and simple for humans to read and for machines to generate and parse files in this Format. It is based on text Format which is completely independent form of language. All of these characteristics mark JSON a superlative data exchange Format. JSON format is described in Fig. 5.

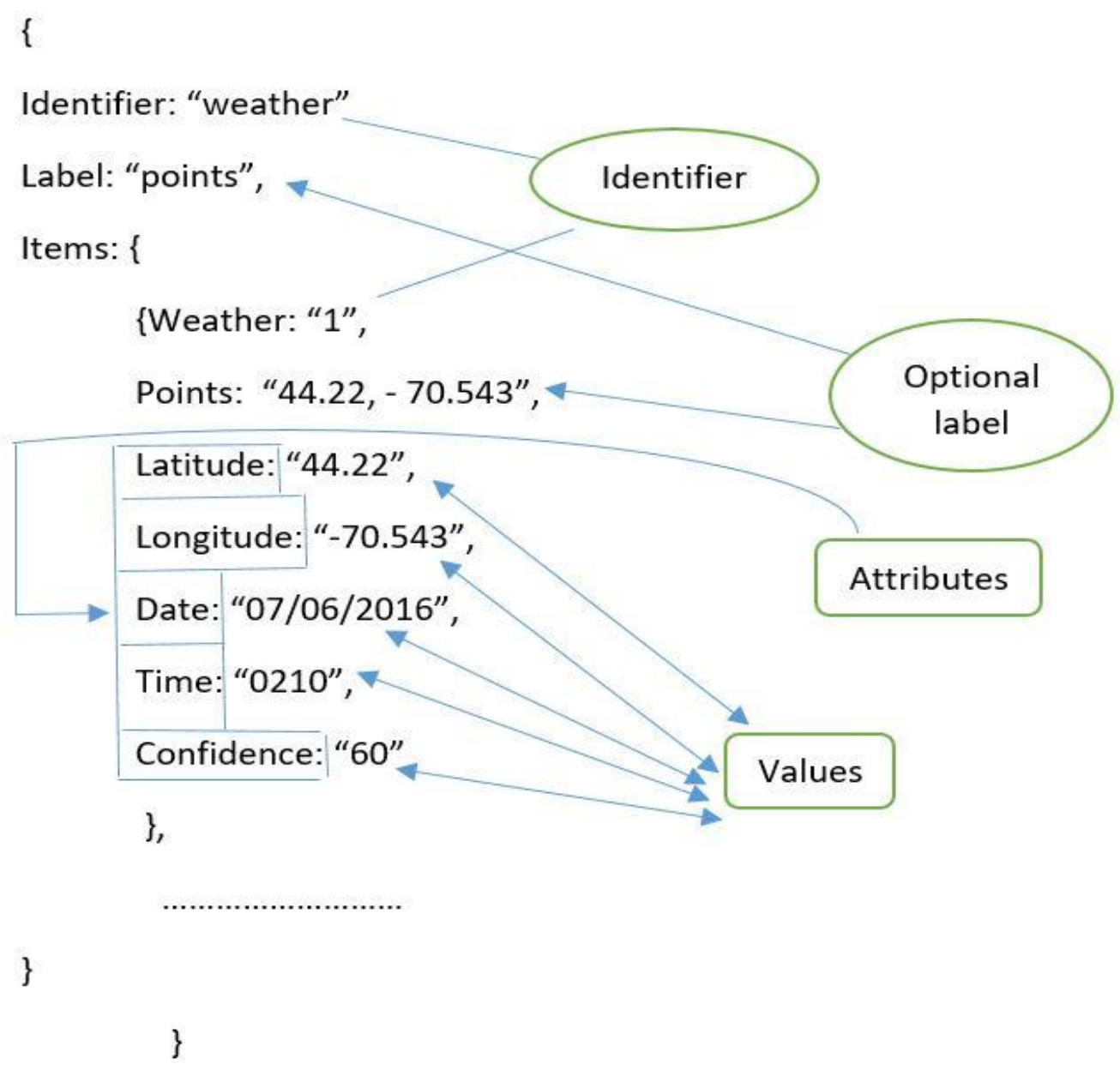

Fig. 5. JSON format

## F. ARFF:

Attribute-Relation File Format (ARFF) contains ASCII coding scheme which illustrates multiple cases that share common features. ARFF have two different units: 1) Header information 2) Data information. The Header section of these files consist of the name of association, an array of traits and types. The ARFF Data unit consist of the affirmation line and concrete instance lines [22]. ARFF is described in Fig. 6.

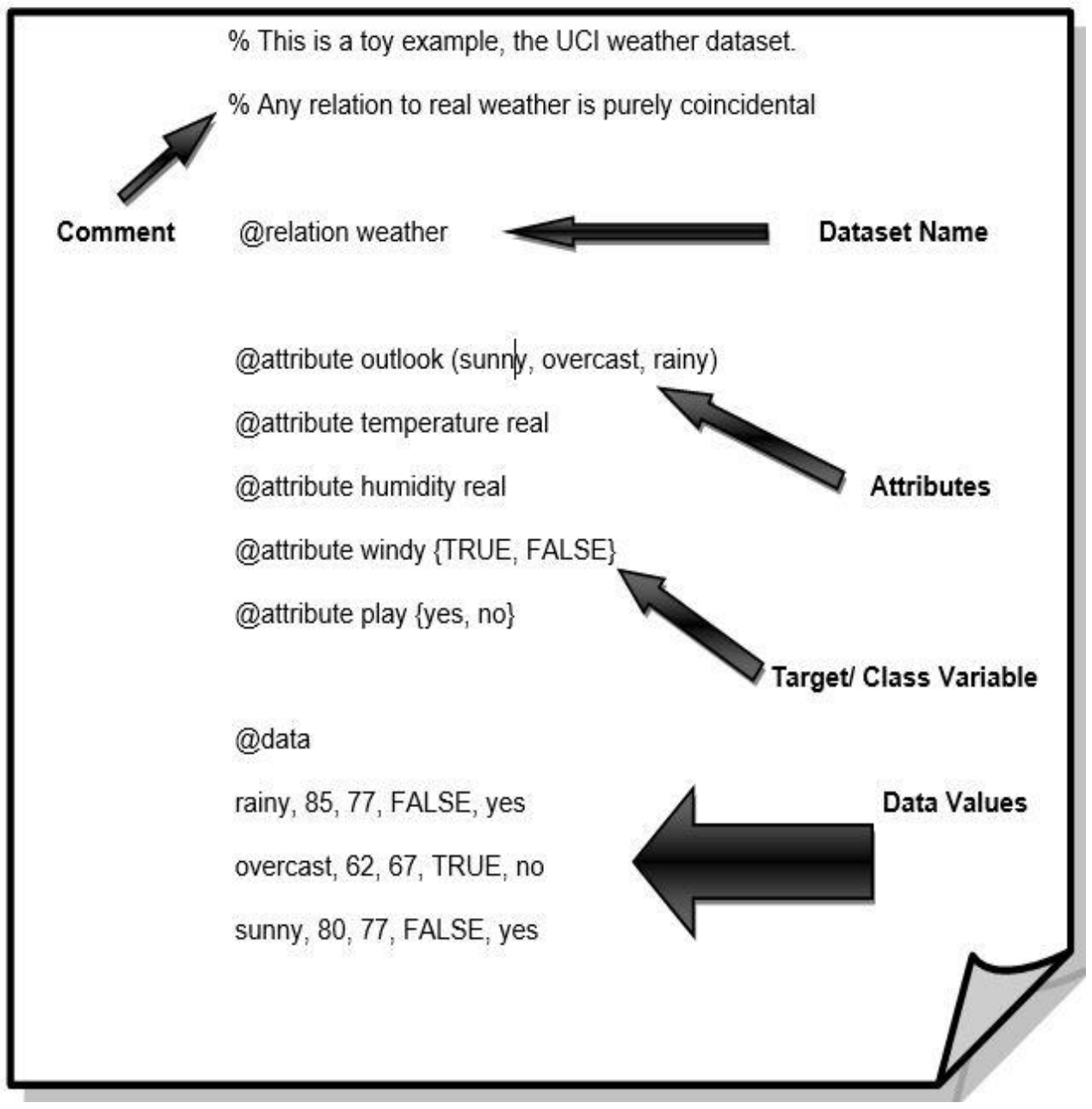

Fig. 6. Attribute-Relation File Format

## G. Sequence File Format:

It is a plane file which consists of binary key/value pairs. Sequence file is mostly used in MapReduce form of input/output file Formats. Temporary output of mapper function in MapReduce paradigm also stored in Sequence file Format. There are three classes available for writing, reading and storing the data. Sequence file format is described in Fig. 7.

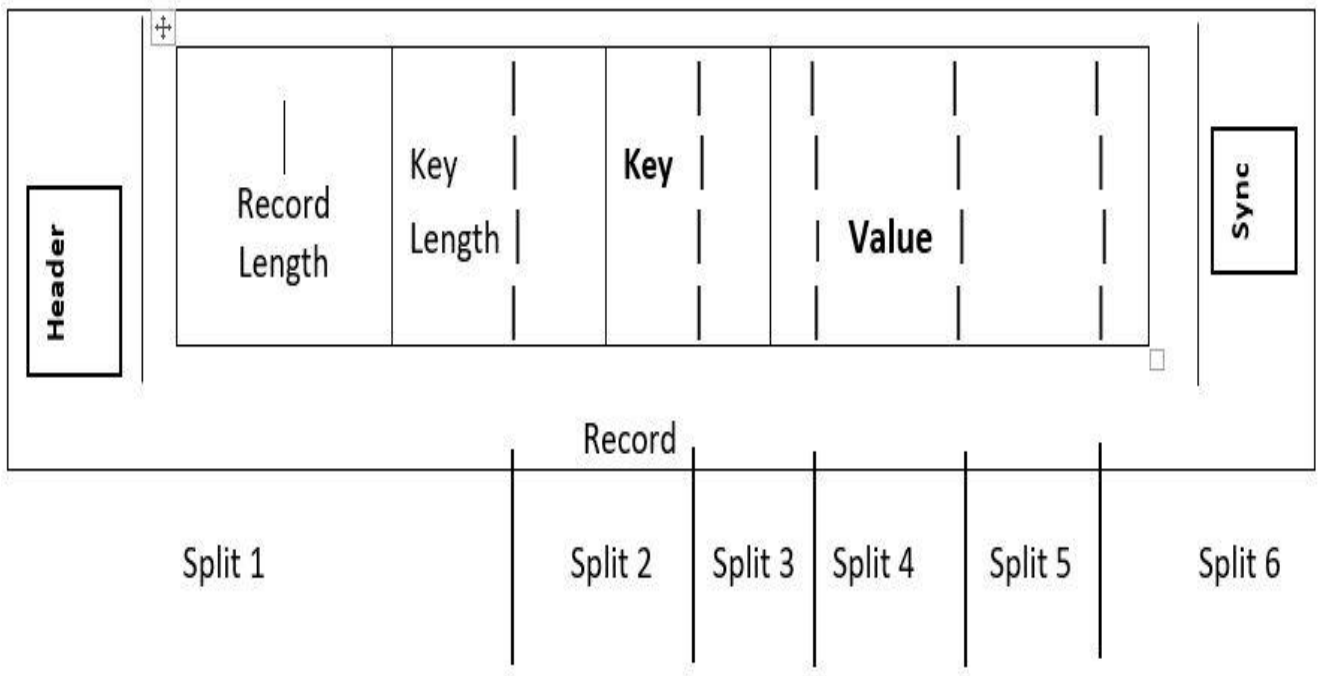

Fig. 7. Sequence File Format

## H. Zip file Format:

Zip file Format is basically only a very thin wrapper around the file Formats which is used to prevent files from splitting. Each and every file in zip file is a separate zip entry and all files in zip file can be unzipped or extracted using its original file name and decompresses the contents of file [23]. Zip file format is described in Fig. 8.

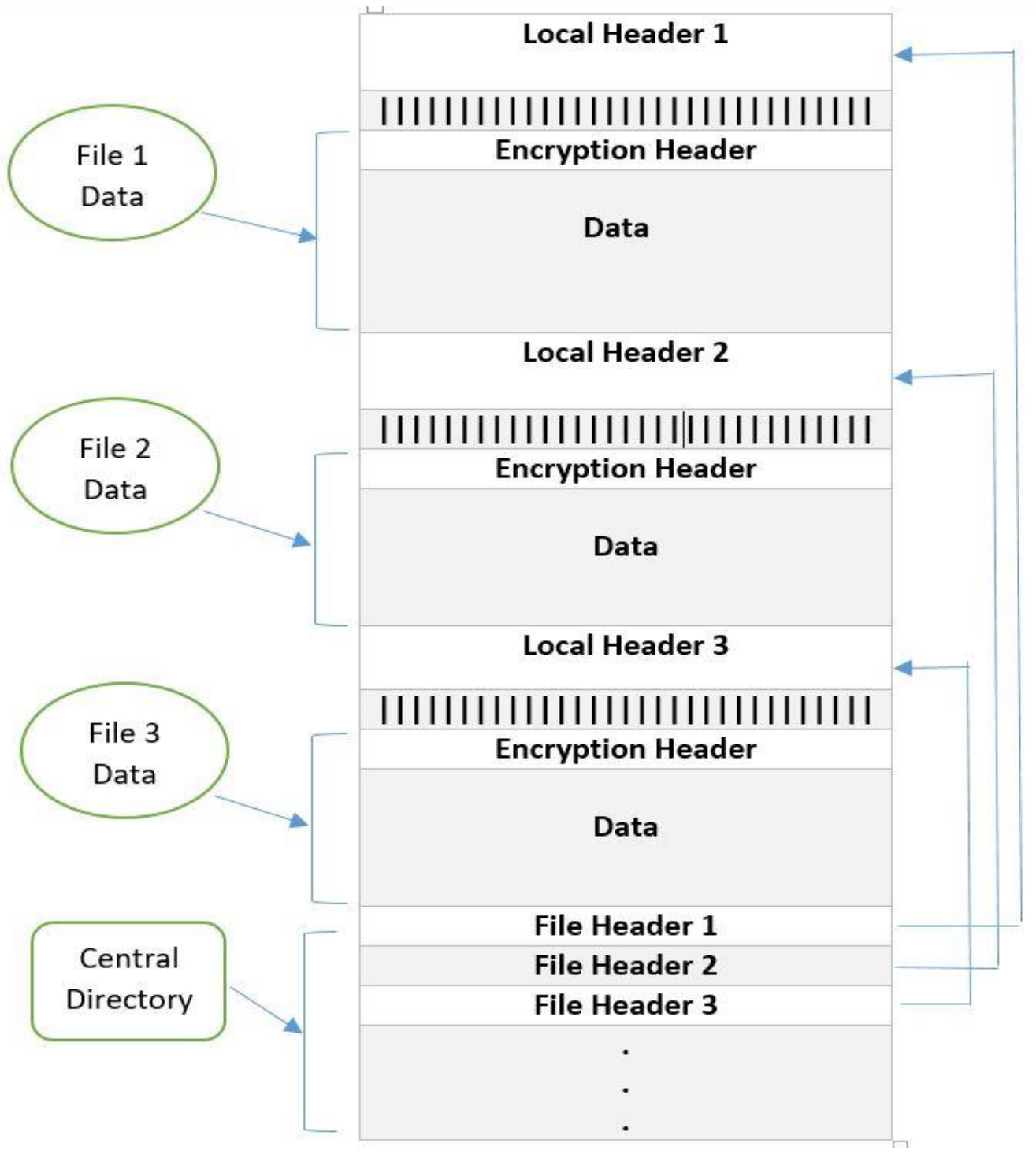

Fig. 8. Zip file format

## I. SQL Database Table:

In SQL database table, data is stored in the form of relations or tables, each and every relation consist of rows and columns. Each and every entry in SQL database table has unique attributes such as entries may in the form of characters, numeric values or dates etc. SQL database tables are used generally to store structured data.

## IV. Modern Data Models (Formats)

A lot of Data Formats exist that are supported by Hadoop Platform for large data storage. Some of these Data Formats are commonly used for general purpose e.g. Text Input Data Format, Sequence File Format, CSV Data Format and JSON Format. Some of them are Modern Data Formats that are supported by Hadoop e.g. Key Value Format, CSR (Compressed Sparse Row) Format, Sparse Vector Format, Binary Format (Avro), LibSVM Format, Columnar Format (Parquet) and ADAM (Avro+Parquet). Data formats used in different machine learning algorithms and techniques for big data analytics are described in TABLE I.

### *A. Key, Value Pair Format:*

In this Format, every text string signifies key and value separated with separator that is used in Hadoop MapReduce Framework for large data storage. In MapReduce, most widely used Format for large Bioinformatics data storage is key value. Hadoop has default data storage Format in Text Input Format. When Machine Learning Algorithms are implemented in MapReduce, mostly used Format is key value. Key value format for input is described in Fig. 9 and key value format for output is described in Fig. 10.

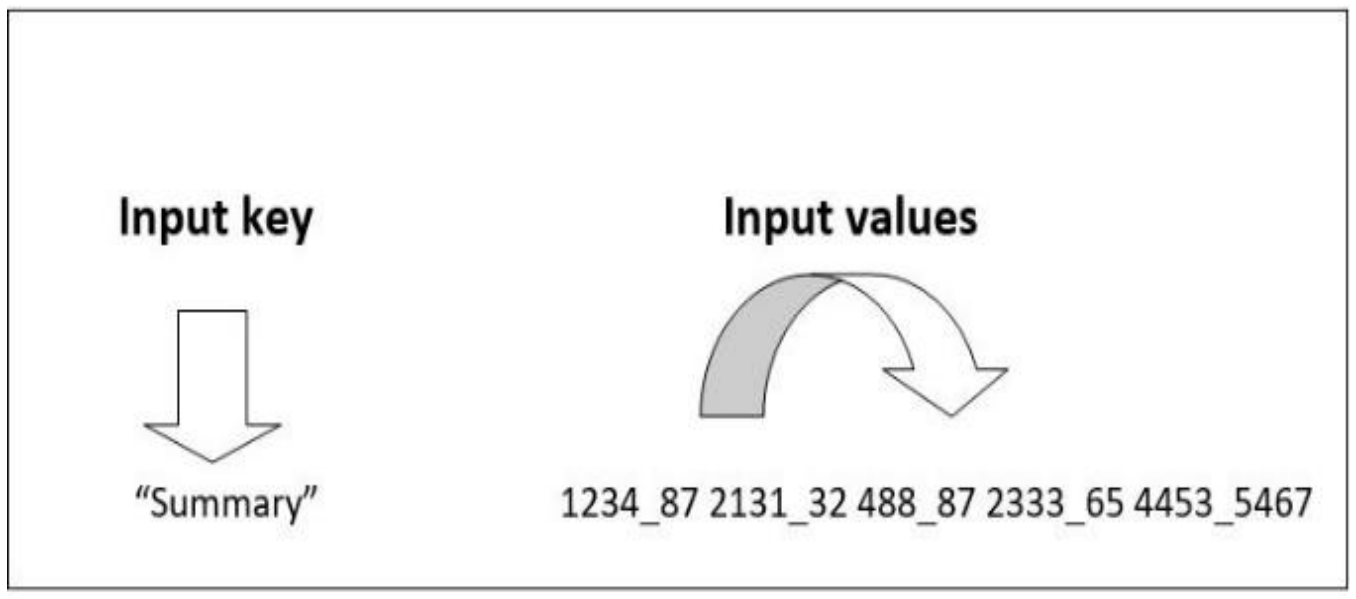

Fig. 9. Input key value pair format

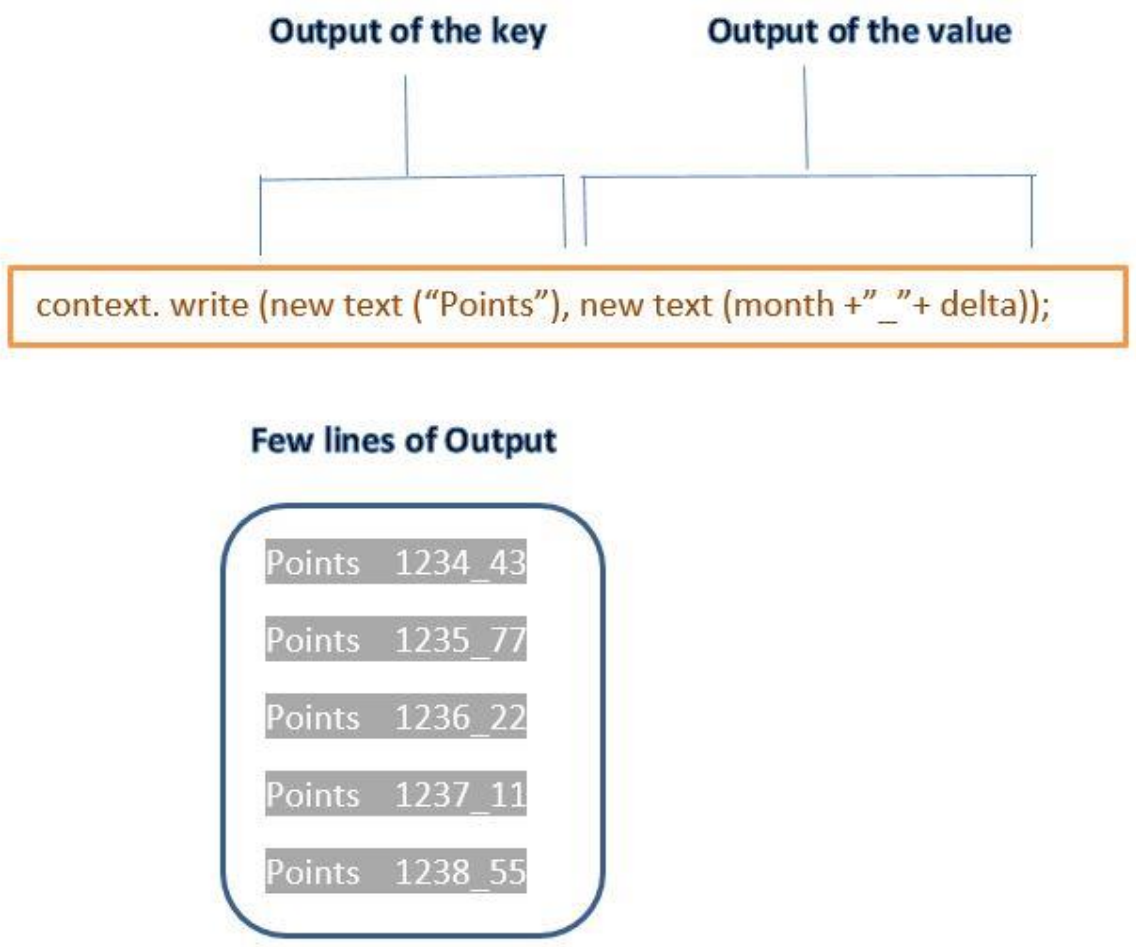

Fig. 10. Output key value pair format

TABLE I. Data Formats Used in Big Data Machine Learning Techniques

| **Machine Learning (Techniques and Algorithms)** | **MapReduce (Mahout Library)** | **Spark (MLib Library)** | **Bioinformatics using Big Data (Bioinformatics+Hadoop)** |
|---|---|---|---|
| NB (Naïve Bayes Bayesian Algorithm) | Text Input Format | LibSVM Format | ADAM (Avro and Parquet) Format |
| GBT (Gradient Boosted Tree Ensemble Algorithm) | (Key, Value) Pair Format | Parquet | - |
| Streaming K-means Clustering | (Key, Value) Pair Format | SparseVector Format | - |
| SVM (Support Vector Machine) | CSR (Compressed Sparse Row) Format | Parquet | ADAM (Avro and Parquet) Format |
| K-means Clustering | SequenceFile Format OR SparseVector Format | LibSVM Format OR SparseVector Format | - |
| Adaptive Model Rules | - | - | - |
| GLM (Generalized Linear Model) | - | CSV Format OR Avro | - |
| LR (Logistic Regression) | (Key, Value) Pair Format | Columnar Format (Parquet) | ADAM (Avro and Parquet) Format |
| Deep Learning | (Key, Value) Pair Format | JavaScript Object Notation (JSON) Format | - |
| Random Forest (Ensemble Algorithm) | (Key, Value) Pair Format | Parquet OR Avro | CSV (Comma Separated Value) Format |
| k-Median Clustering | (Key, Value) Pair Format | SparseVector Format | - |
| kNN (Instance based Algorithm) | BinaryFormat (Avro) OR (Key, Value) Pair Format | CSV Format | ADAM (Avro and Parquet) Format |
| Association Rule Mining (Apriori Algorithm) | (Key, Value) Pair Format | LibSVM Format | Vertical data layout Format |
| Decision Tree | CSV (Comma Separated Value) Format | LibSVM Format | - |
| Linear Regression | (Key, Value) Pair Format | LibSVM Format | ADAM (Avro and Parquet) Format |

### *B. CSR (Compressed Sparse Row) or CSR (Compressed Row Storage) or Yale Format:*

It includes matrix that consists of 3 one-D (one Dimensional) arrays of non-zero elements (dense Matrix) in the form of rows. In Compressed Column Storage (CCS) Format, Columns stored in the Matrix instead of rows. When SVM Algorithm is implemented in MapReduce Framework for large Datasets, CSR or CCS Format is used. CSR format is described in Fig. 11.

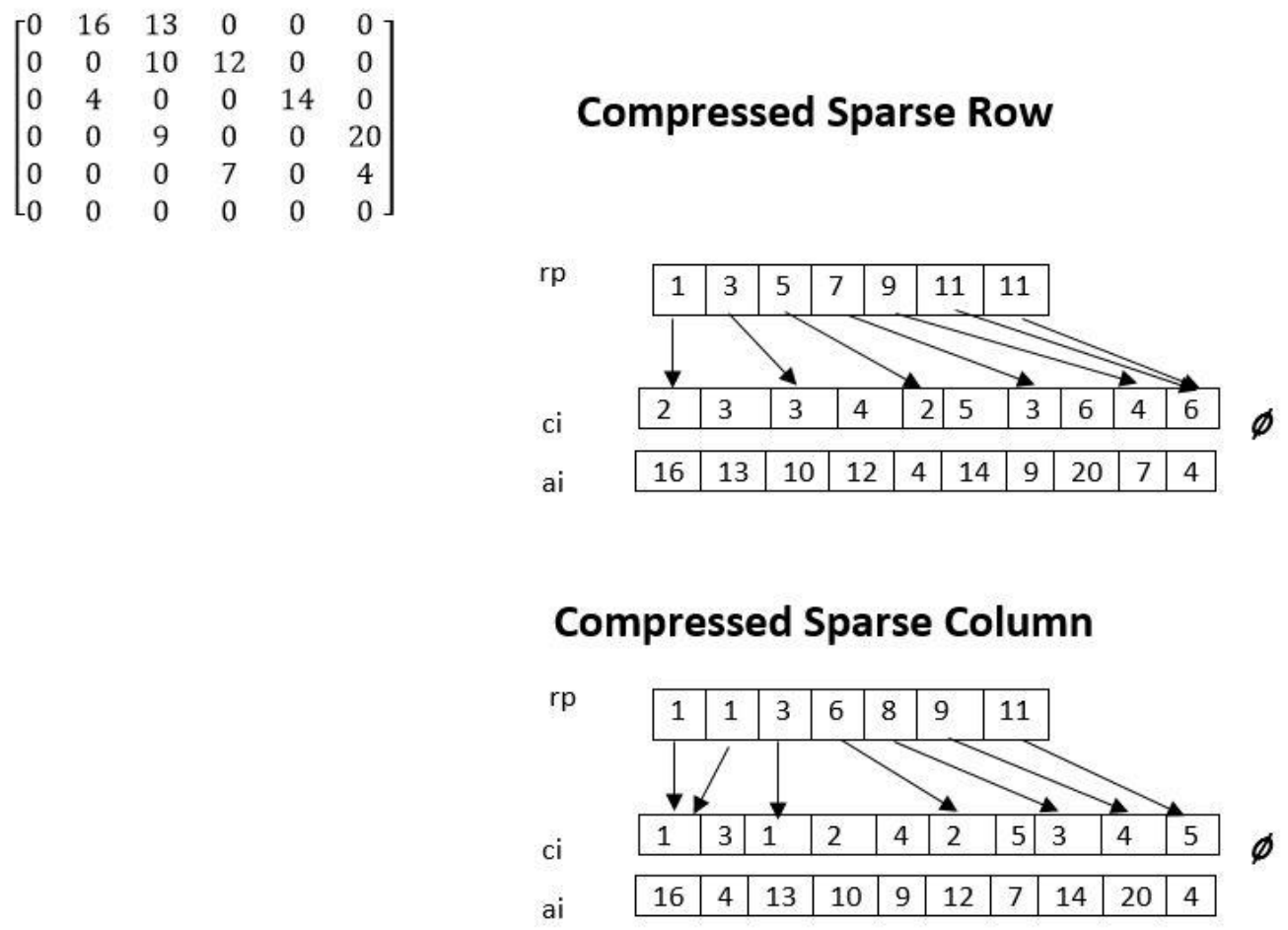

Fig. 11. Compressed Sparse Row

### *C. Sparse Vector Format:*

It includes one-D (one-Dimensional) array of zero elements (Sparse Matrix). Elements of Sparse Vector are represented by Linked list. We can create Sparse and Dense Vectors in Spark Framework using MLib (Machine Learning Library). Sparse Vector Format is generally used for storage of Clustering data. Sparse vector format described in Fig. 12.

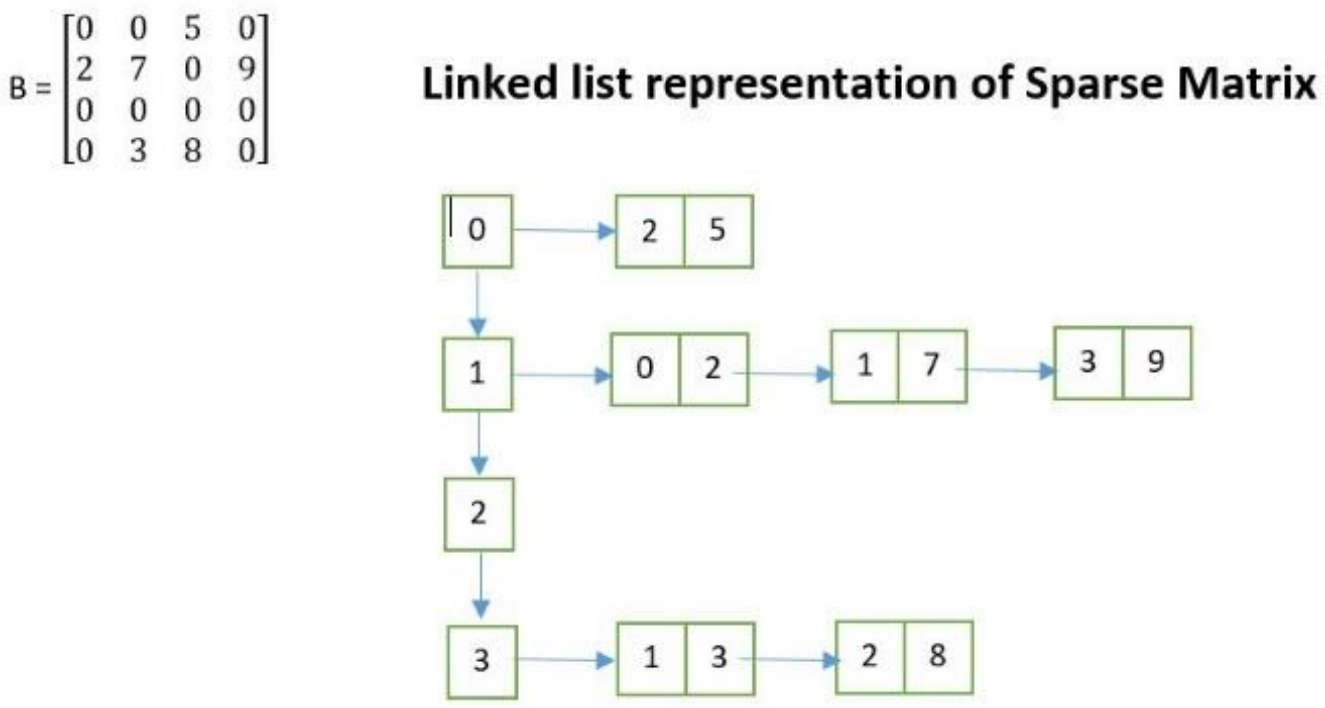

Fig. 12. Sparse vector format

### *D. LibSVM Format:*

It is used by MLib (Machine Learning Library) in Apache Spark. This Format is used for Classification and Regression purpose for the storage of sparse data such a way that non-zero values are involved in large Dataset. In this Format, every line consist of an instance and it is end with '\n' character. Label consist of target value for Regression. It is supported by multi-class, not for one class SVM.

It is helpful for sparse training data. In this Format, first value of first line describes the dataset label. Then, index is representing after blank space. After the colon, value is given of specific index. In each line, specific label is given for dataset.

It is open source library that is used in ML Classification, Clustering and Regression. It compresses the dataset size. Reducing the size of dataset is very important for improving better analysis. It is efficient for both binary and multi-class Classification. Main feature of libSVM is that to provide GUI (Graphical User Interface). libSVM package includes C++ and JAVA source code library. File extensions are zip or tar.gz file for libSVM data storage Format. LibSVM format is described in Fig. 13

```
-1 4:1 19:1 21:1 30:1 38:1 41:1 54:1 65:1 69:1 74:1 76:1 79:1 80:1 85:1
-1 5:1 9:1 18:1 24:1 28:1 32:1 41:1 55:1 66:1 75:1 82:1 88:1 91:1 105:1
-1 5:1 19:1 22:1 31:1 43:1 46:1 52:1 58:1 66:1 67:1 77:1 82:1 87:1 99:1
-1 6:1 7:1 16:1 21:1 37:1 44:1 48:1 65:1 68:1 71:1 78:1 79:1 81:1 85:1
+1 7:1 12:1 14:1 33:1 38:1 41:1 56:1 66:1 69:1 77:1 79:1 80:1 91:1 99:1
```

Fig. 13. LibSVM format

### *E. Avro Format:*

It is row based binary serialization Format. It represent encodings such as Binary and JSON (for web applications) [24] [25]. Avro depends on schema and frequently used Format in Apache Spark for large Bioinformatics (Genome) data storage. Avro integrated with Dynamic languages and support for Dynamic Typing. Avro format described in Fig. 14.

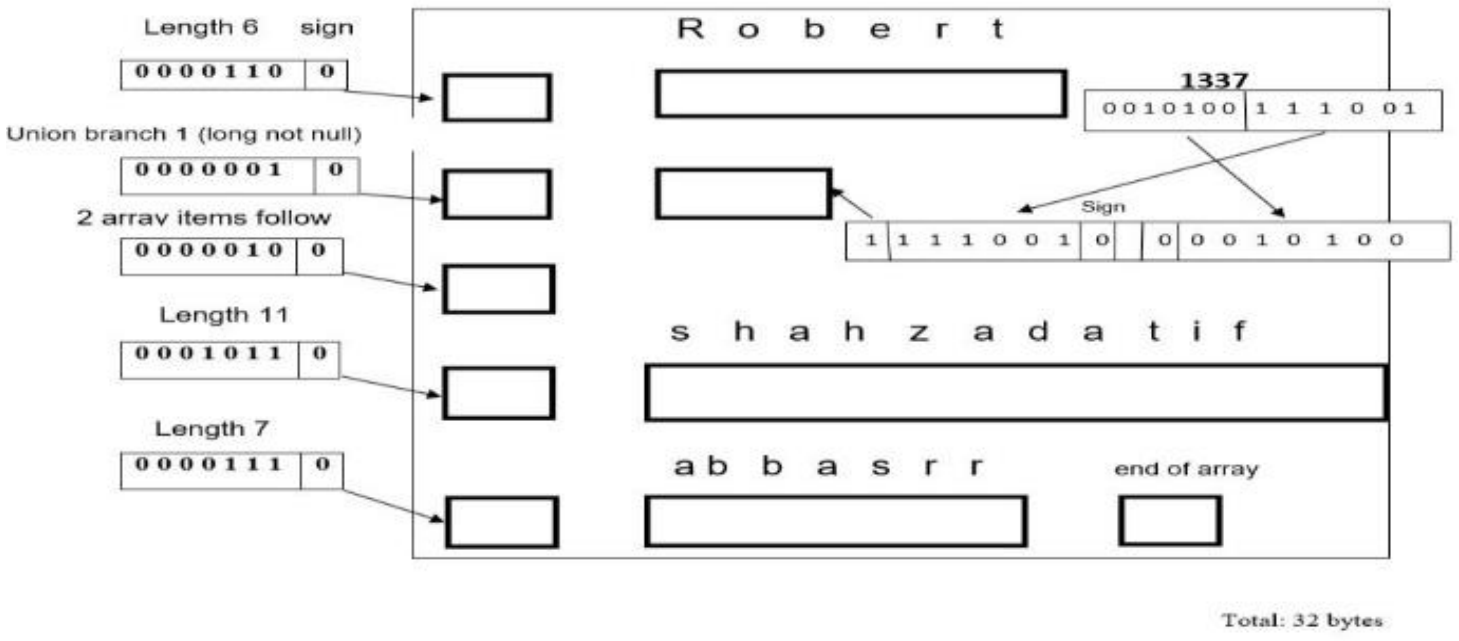

Fig. 14. Avro format

### *F. Parquet format is columnar data storage Format:*

Parquet is columnar data storage Format: In this Format, Binary data stored in the form of columns. It is very effective and provides best query Performance on Hadoop platform. Like Avro Format, Parquet is supported by Apache Spark Framework for the storage of large Bioinformatics data. Parquet format is described in Fig. 15.

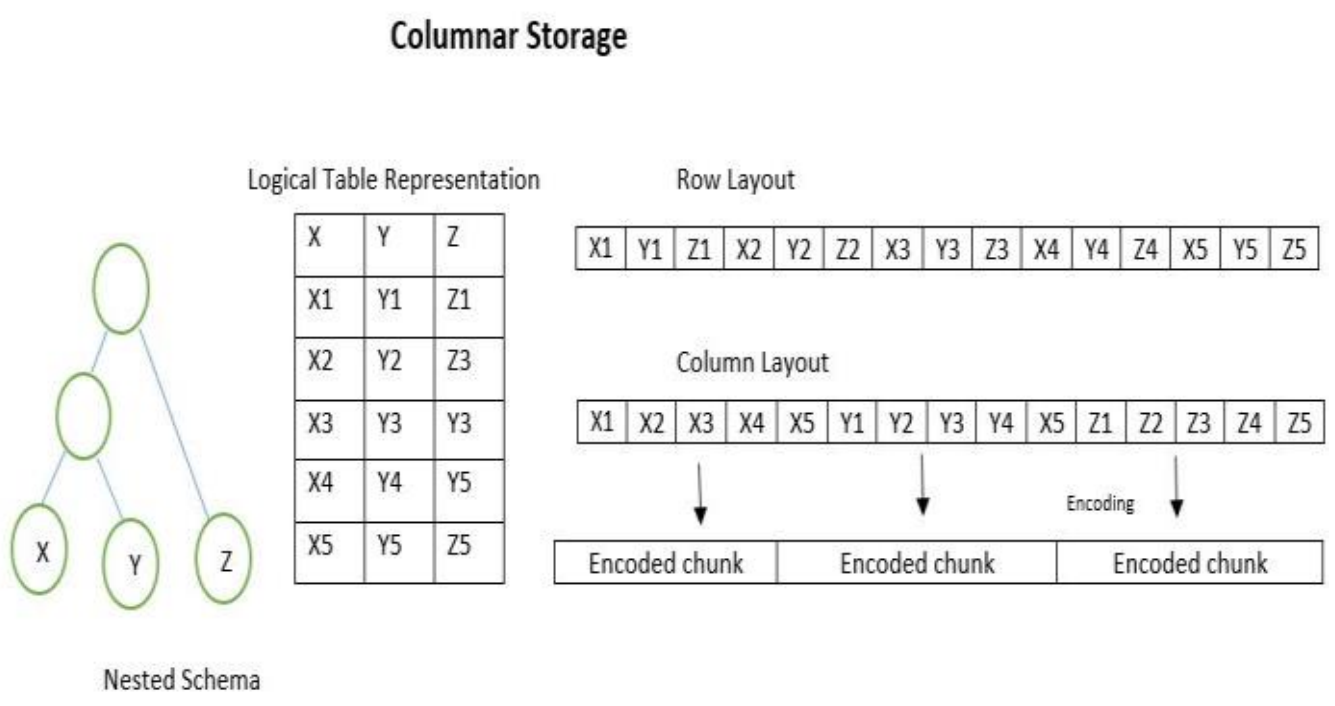

Fig. 15. Parquet format

### G. ORC (Optimized Row Compressed) Format:

It is used for relational data storage in Hive Framework that consumes minimum space. It performs better without compression than other Data Formats in Hive. It consumes less execution time for reading, writing and accessing Hive data. ORC format is described in Fig. 16.

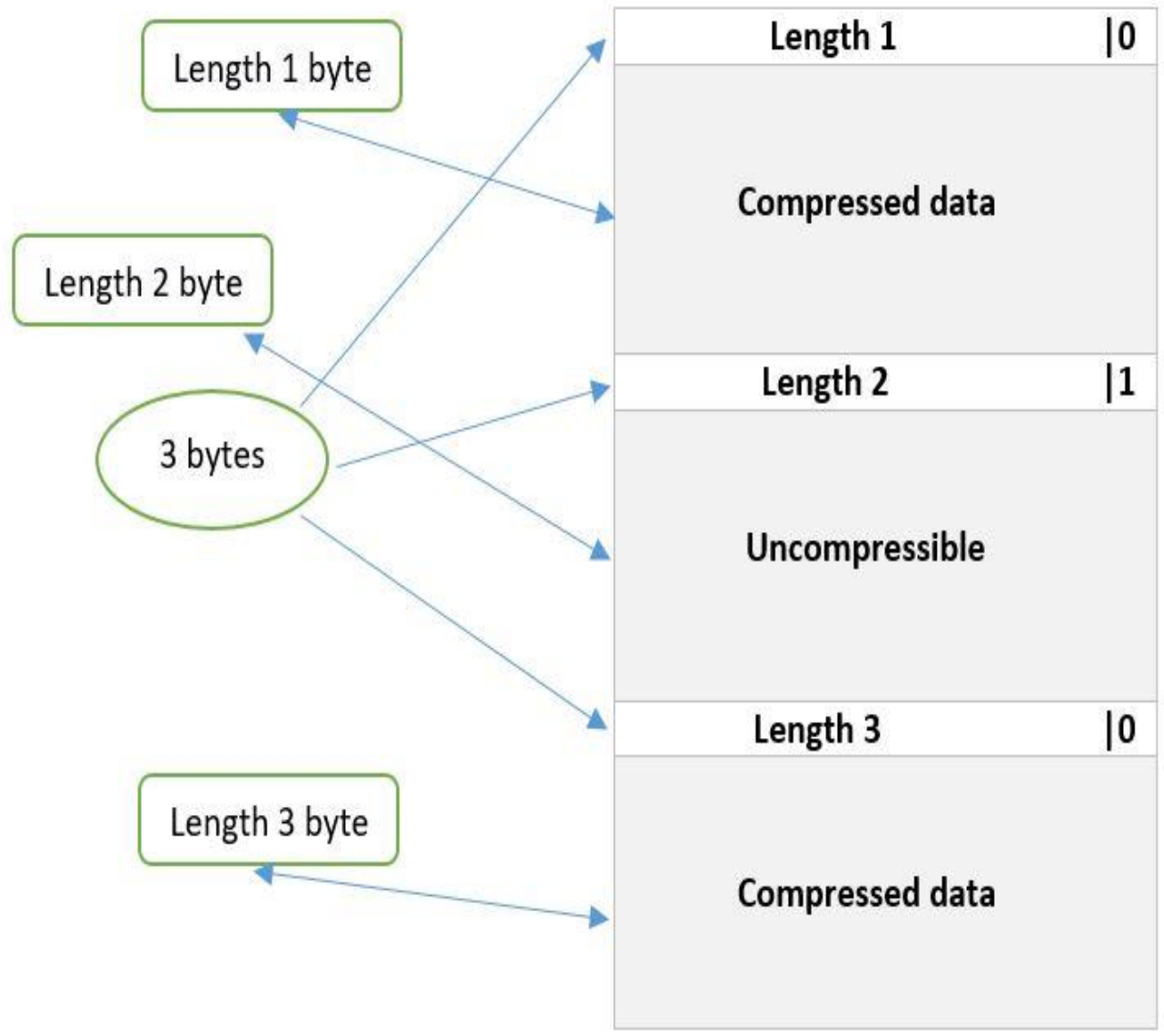

Fig. 16. Optimized Row Compressed format

### H. BAM (Binary Alignment Map) File Format:

It is used for the storage of text data into Sequence (Aligned or Unaligned) for Genome in single node processing for direct access [26]. Data cannot be stored in HDFS without accessing BAM Library. When data is stored on Hadoop platform, Spark queries are used for sequencing data [27]. BAM file format is described in Figure. 17.

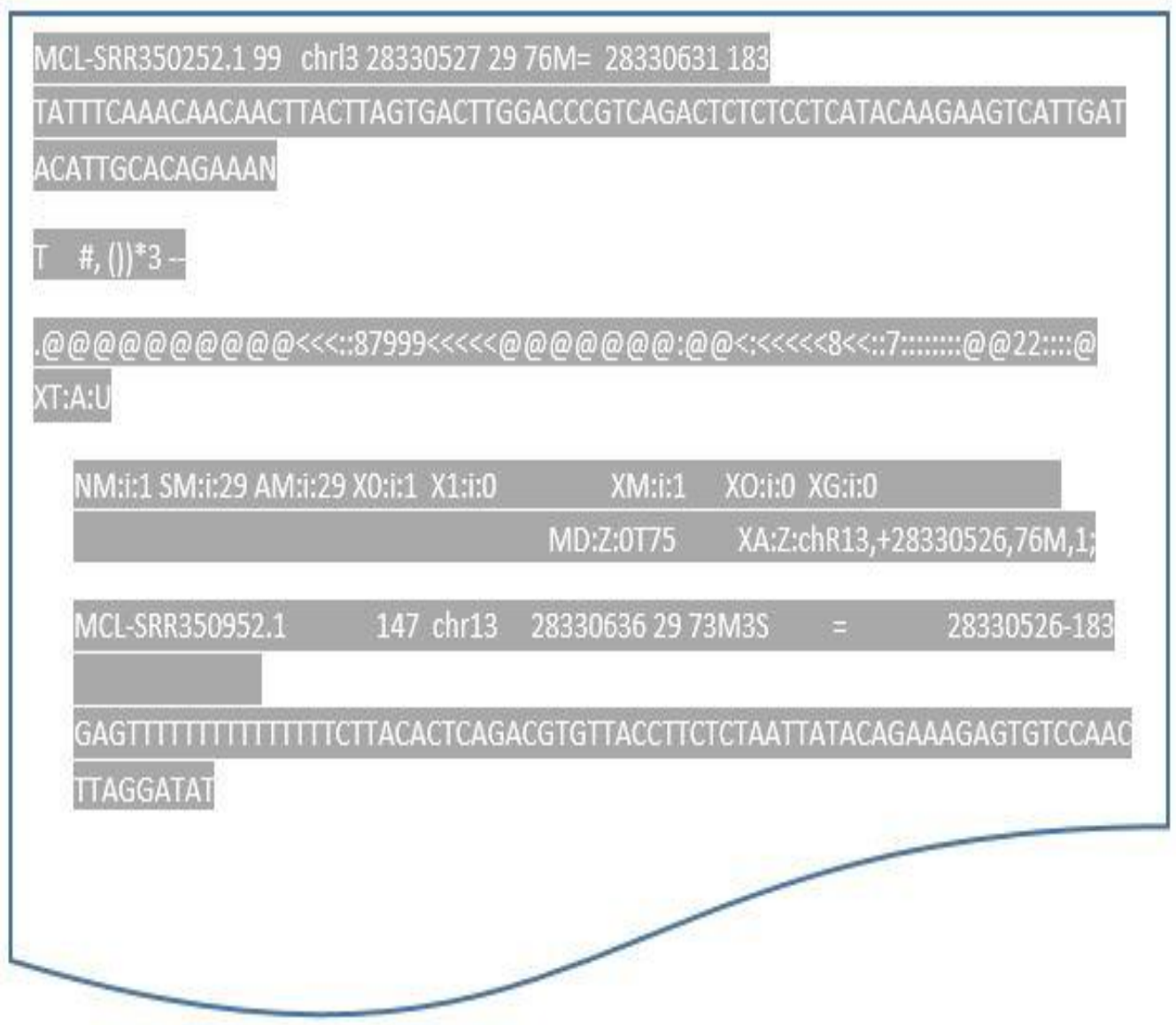

Fig. 17. Binary Alignment Map format

### I. Fastq Format:

It is text Format for storage of DNA, RNA and Genome sequence data with their quality. Fastq format is described in Fig. 18.

It is used in multiple and sequence alignment for Bioinformatics data storage. In this dataset Format, first line represents @ character that is sequence identifier, second line starts with sequence letters, third line starts with + character, fourth line describes quality scores for sequence letters given in second line. Quality score based on ASCII (American Standard Code for Information Interchange). This format helpful for comparing multiple sequences. Fastq compression files are in GNU zip format. Compression of quality scores is very significant for reducing storage requirements. File extensions for Fastq data storage are .fq and .fastq.

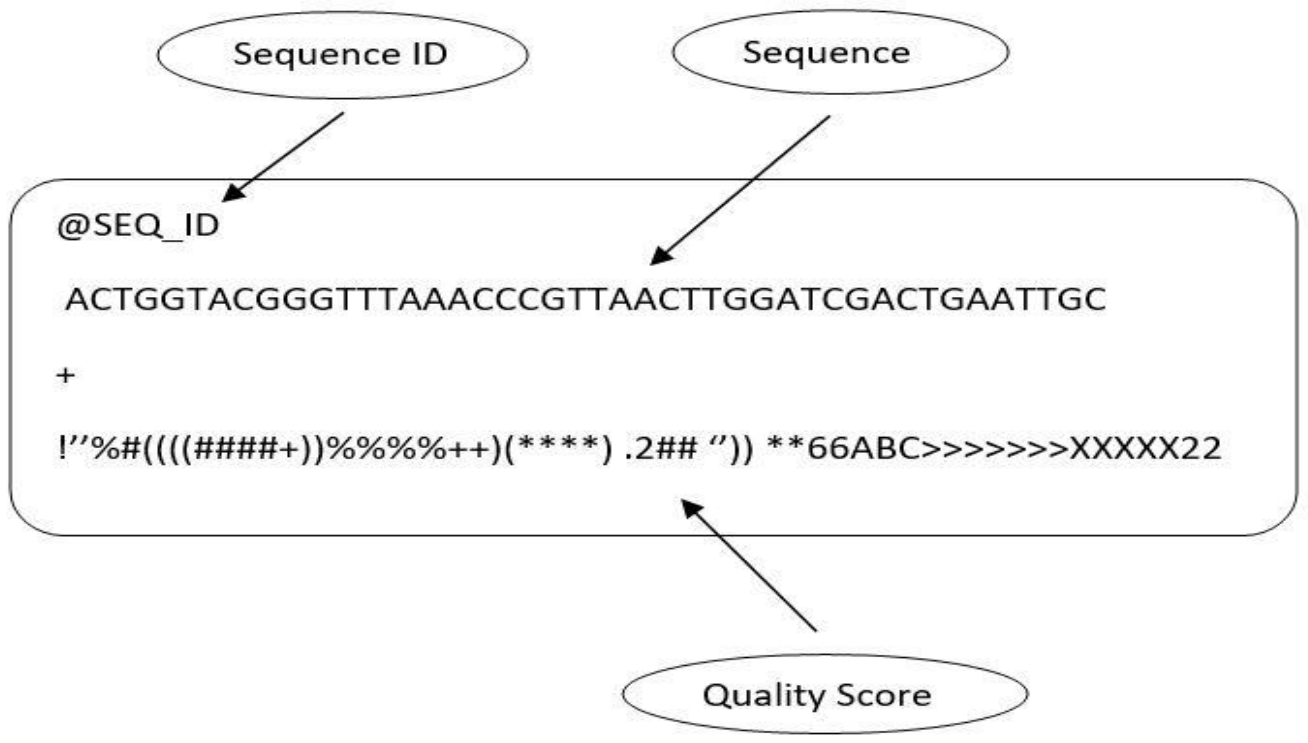

Fig. 18. Fastq format

### *J. FASTA Format:*

It is store the Nucleotide Sequence Biological information. In this Format, multiple sequences exist. This Format starts from > symbol that determines the sequence name and output the sequence according to this name. FASTA format is described in Fig. 19.

In this Format, first line represents > symbol or ; (semicolon) depicted header line that is sequence identifier, second line starts with sequence letters. Length of each line sequence less than 80 characters. Sequences are described by amino acid codes. Compression is very important for FASTA like Fastq. File extensions for Fastq data storage are fasta, fna, ffn and frn. Blank spaces are not all permitted in the middle of sequence.

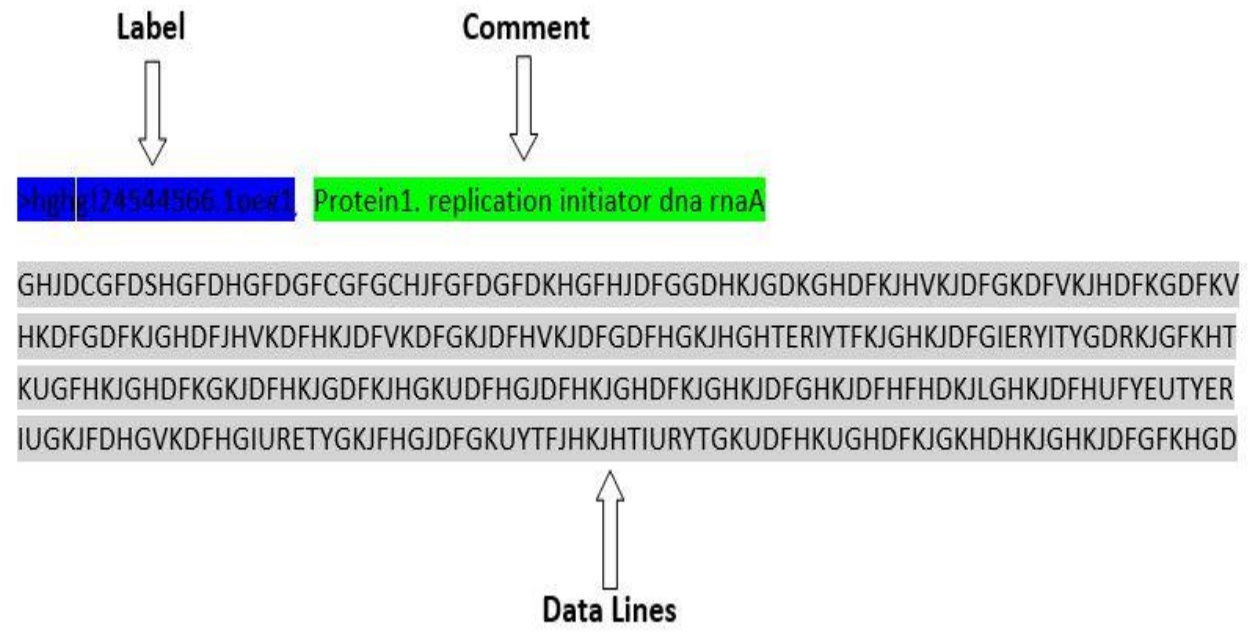

Fig. 19. Fasta format

### *K. VCF (Variant Call Format):*

It is designed for the storage of large Genomics data in the form of text including some special keyword # in MapReduce or Spark Framework. In VCF, every line is represented in the array. VCF provides better Speed when large genome data is stored in Spark instead of MapReduce. Spark Transformations are performed directly from VCF. VCF is described in Fig. 20.

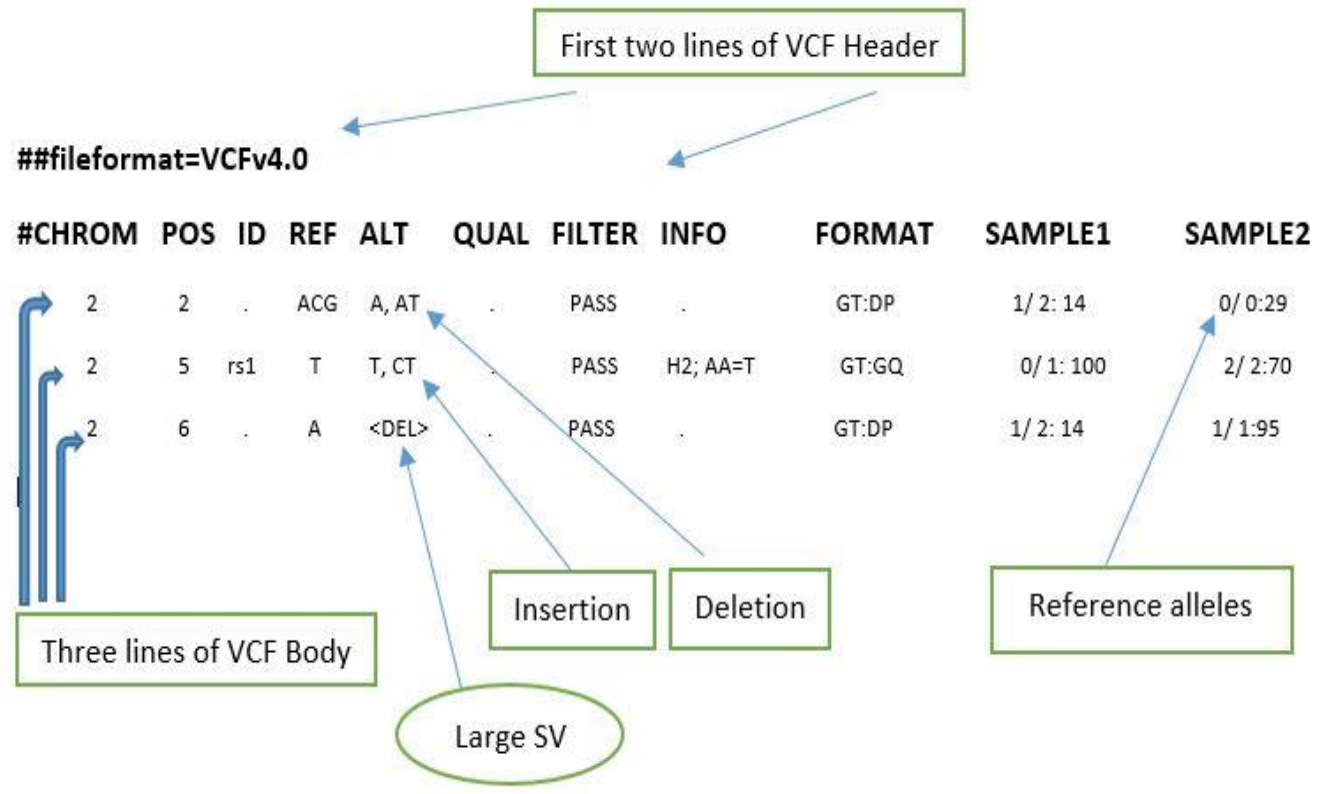

Fig. 20. Variant Call Format

### *L. Graph Format:*

Giraph packages include input Formats such as Vertex Input Format (for Directed edges) and Edge Input Format (for Relational storage) [28]. Their output Formats are Vertex Output Format (for representing writing data for every vertex) and Edge Output Format (for representing writing data for every edge) [28].

### *M. Audio and video Format:*

When video comes from medical and clinical data in Hadoop platform that data is converted in to MPEG-2 Format which is specially design to perform complex tasks on videos through video Transcoders. This MPEG-2 Format converts that video Format into specific sequence file Format that take binary key value pairs to Hadoop platform through Mappers for better results and get better performance [29] [30]. Like video Formats, audio data first converts into MPEG-2 through transcoders and then that MPEG-2 converts to sequence file to perform big data tasks on Audio files.

## V. TOOLS FOR DATA FORMAT CONVERSION

When large Bioinformatics data is stored in Hadoop MapReduce or Apache Spark Framework, then this data will be stored in one specific Hadoop supported Format. Some Data Formats are not supported by Hadoop Platform. To reduce this problem, conversion tools are used that convert Data Format from one form to another. Many conversion tools are used that convert Bioinformatics Data Formats into Spark supported Data Formats. Most important conversion tool is TCS that transform data from Relational Databases to Hadoop Platform. When Machine Learning Algorithms and Techniques are implemented in Apache Spark, Avro and Parquet are supported Formats. In this situation, ADAM stack plays a significant role in Spark Framework.

There are many tools that are used for conversion of large dataset from one format to another. Some tools are Data Converters, FME data conversion and integration tool, Altova MapForce, SQL database table, TCS, Sqoop, Avro2parquet, Apache Drill, Format Converter, ADAM, Apache Giraph and GraphChi.

### *A. Data Convertors:*

Data convertors is a Python Library and a command line tool that is used to perform routine data conversion tasks easily. Data convertors help data scientists to perform commonly tasks very efficiently and easily for example moving data between tabular Data Formats, converting a csv file to JSON object or an Excel spread sheet to csv [31].

### *B. FME data conversion and integration tool:*

FME is the only spot-on spatial ETL (extract, transform and load) platform that is used to cope with the spectrum of data interoperability challenges completely and easily, including proprietary managing and evolving Data Formats. FME data converters also help in restructuring, adapting to new schemas, distributing and integrating the data.

With the help of the power and flexibility of FME platform, it is easy and quick to enable spatial data translation and migration projects, exchange spatial data, distribute data to multiple users and transform data from one Data Format / model to another. FME data converter supports hundreds of Formats and data validation too [32].

*C. Altova MapForce:*

MapForce is a multi-purpose IDE that is used to transform data from one schema to another schema or from one Format to a number of file-based Formats or other specific Format without writing any program code, conveniently and easily. MapForce determines the structure of data or gives an option to provide the schema of data from user. MapForce supports a lot of variety of Data Formats to work with. MapForce also gives graphical view of the data to the user for mapping transformations among different Data Format files [33] [34].

*D. SQL database table:*

Most of Machine Learning tools takes different types of Data Format as input. In most cases, the dataset is not existing in a single file and dispersed across diverse sources such as CSV, Text and TSV etc. For combining all of the available Formats in a single Format, a tool is required to collect, filter, intersect and transform data. When data is large in size and the modifications are multifaceted such as combination of quite a lot of sources, purifying a subsection of dataset or dealing with a huge number of rackets or rows, a more dominant tool like RDBMS is required. MYSQL is an excessive tool to organized data into one Format that are sparse across very different sources or Formats [35].

*E. TCS:*

It is secure data migration tool that migrate large data from Relational Databases and File Systems to Hadoop platform such as HDFS, HBase and Hive. This tool is cost effective and efficiently transform data from one form (multiple sources) to another form [36].

*F. Sqoop:*

It is import and export Hadoop tool that migrate large data from Relational Databases to Hadoop platform in HDFS, HBase and Hive. It imports data into HBase and Hive [37]. Sqoop does not support for Parquet Format export but support for Avro Format export.

*G. Avro2parquet:*

It is a data conversion tool that is based on Hadoop MapReduce Framework that easily convert data from Avro (Row-Oriented form) Format into Parquet (columnar form) Format. This tool works in the form of key value pair such as (GenericRecord, NullWritable) in Avro container [38].

*H. Apache Drill*

It is a SQL (Structured Query Language) query engine that supports HDFS, HBase, MongoDB and Amazon S3 on Hadoop platform and also include many Built-in-Functions for many data type conversions [39]. Drill is used when CSV or TSV (Tab Separated Value) files are converted into Parquet Format because Parquet offers best Performance than CSV or TSV test files [40].

For converting CSV files into Parquet Format in Spark Framework, first of all spark-csv package is used to load CSV data in Spark data frames, then these frames are saved to Parquet Format [41]. In Spark Framework, MLib (Machine Learning Library) contain Utilities such as 'MLUtils' that are used to convert matrix and vector columns into new spark matrix and vectors.

*I. Format Converter:*

It is a tool that convert the DNA, RNA and Proteins Sequences into specified Format for Bioinformatics data. In this tool, we give the input and output Formats and also gives our sequence. For example, FASTA is converted to CSV Format.

*J. ADAM:*

It is open source protocol stack that includes API's and set of Formats used in Apache Spark Framework for processing and storage of large Genomics data [42]. ADAM is implemented on top of Parquet (for accessing database such as Impala) and Avro (for explicit schema access in any language such as Python, Scala, C++ and Java) [42].Existing Formats such as BAM/SAM/VCF are not scale with large Bioinformatics data due to limited Scalability using BAM and deficiency of explicit schema [42].

*K. Apache Giraph:*

It is an open source system on the top of Hadoop I/O Format API for large processing of graphs on Hadoop Platform. It automatically includes existing Formats of Hadoop.

*L. GraphChi:*

It is a system for large processing of graphs in which ML Algorithms are executed on large graph only on single computer [GraphChi paper ref]. It reduces the several problems for graph processing and storage that occurs in Hadoop Platform. It takes less Execution time than the time taken by the Hadoop and Spark for graph processing [43]. GraphX is a built-in Library for large and faster graph processing in Spark Framework.

## VI. Discussion

There are many Modern Data Models (Format) that are supported by Hadoop and Apache Spark for applying Machine Learning Algorithms on large Bioinformatics data. It is very important to choose a specific Data Format that is suitable for efficient processing and storage of huge Bioinformatics data.

The objective is to select a data format that takes less space and provides great performance. The comparison of Performance and Speed of different data formats is described in TABLE II.

Now we will discuss about Performance of existing Data Models that are used on Hadoop or Spark Framework for applying Machine Learning Algorithms on large Bioinformatics data. Sequence File Format provides better performance for faster datasets than using CSV or Text Formats. Supported platforms for Sequence File Format are MapReduce and Hive but it does not perform well for large datasets. MapReduce and Hive provide support for CSV, Text and JSON Formats, whereas Weka provides support for ARFF.

Avro Serialization Format provides best performance when we have data of multiple Rows instead of Columns. It is less efficient than Parquet Format and ADAM stack. Better speed has been achieved using Parquet Format than Avro (Serialization) Format in Hadoop and Spark Framework [44]. In case the data contains many columns then Parquet or ORC Format should be used that provide best speed and performance. If we have many rows, then Avro is best Choice [44]. Actually, Parquet compress data that has advantages of save the space. Supported Platforms for Avro and Parquet Formats are Hadoop MapReduce, HBase, Hive, Pig and Spark. ORC Format supports for indexing and better for query processing. Supported Platforms for ORC Format are Hadoop MapReduce, Hive, Pig and Spark but it is not supported by HBase.

Machine Learning Classification and Clustering play an imperative role in Bioinformatics. Microarray (Genomics) data will be classify into binary and multi-classes with the help of ML Classifier. K-means and k-median Clustering are very beneficial for Bioinformatics data. Better analysis will be perform by using ML classifiers in Apache Spark and Hadoop MapReduce framework for Bioinformatics data. Apache Spark use mlib library and MapReduce use mahout library for analysis of Bioinformatics data.

For application of Machine Learning Algorithms and Techniques in Hadoop or Spark, LibSVM Format and ADAM Tool are frequently used for large Bioinformatics (Genomics) data. LibSVM plays an important role in ML Classification and Clustering. ADAM takes great benefit from Parquet and Avro by using Apache Spark in the application of ML Algorithms for large data Genomics. ADAM has better performance and execution time (Speed) than traditional BAM/SAM or VCF for large Bioinformatics data. Supported platform for ADAM is Apache Spark.

Following guidelines may be used for choosing a specific data format:

- If multiple Queries are running in Impala, then Parquet Format will be used and ORC (Optimized Row Columnar) Format will not be used.
- If running Multiple Hadoop MapReduce tasks in pipelining, then Sequence File Format will be used.
- If running large Queries in Hadoop Platform, then Apache Parquet (columnar) Format is used.

TABLE II. PERFORMANCE COMPARISON OF DIFFERENT DATA MODELS (FORMATS)

| **Data Models (Formats)** | **Performance** | **Speed** | **Supported Platforms** |
|---|---|---|---|
| CSV | Better than JSON file format | Less than Sequence file format | MapReduce/Hive |
| Text | Less than CSV format | Less than Sequence file format | MapReduce/Hive |
| Sequence | Better for small datasets | Faster for small datasets but not for large datasets | MapReduce/Hive |
| JSON | Less than CSV format | Less than CSV format | MapReduce/Hive |
| ARFF | Better than CSV format | Less than Sequence file format | Weka |
| Avro (Serialization) | Less than Parquet and ADAM Stack.<br>Best for multiple Rows. | Less than Parquet.<br>Less Scalable for many Columns. | Hadoop MapReduce, HBase, Hive, Pig and Spark. |
| Parquet | Better than Avro.<br>Best for multiple Columns.<br>Less than ADAM Stack.<br>Not support for indexing. | Better than Avro.<br>Less Scalable for many Rows. | Hadoop MapReduce, HBase, Hive, Pig and Spark. |
| ORC (Optimized Row Compressed) | Better than Avro.<br>Support for indexing<br>Best with Hive | Better than Avro.<br>Best for Querying processing. | Hadoop MapReduce, Hive, Pig and Spark.<br>Not Support for HBase |
| ADAM Stack | Better than simple Avro, Parquet and VCF. | Better than simple Avro, Parquet and VCF. | Best support for Apache Spark. |

## VII. CONCLUSION

Bioinformatics research involves large volumes of data and complex data analytics. Most of the tools in bioinformatics use iterative machine learning methods. These tools can be scaled to handle large volume of data by using parallel and distributed computing models as provided in Hadoop and Spark. Big data tools and platform use different data models and formats than used by traditional bioinformatics tools.

Data model or format plays an important role in data analytics. It affects the understanding of data, representation of data, space required to store data, data I/O during processing of data, application of machine learning or mining algorithm on data, intermediate results of processing, in-memory analysis of data and overall time required to process data. So, a careful selection of data model or format is required for conducting any data analysis experiment.

In this paper, state of the art data models and formats are explored. Each format is elaborated with its salient features and storage mechanism. Particularly, data models for big data platform are presented that can be used to conduct bioinformatics experiments. Data models are also discussed in relation to implementation and application of machine learning algorithms. Finally, modern data sets are compared on the basis of their performance on big data platform.

This study will help researchers and data scientists in getting firsthand knowledge of state of the art data formats and in better planning and accomplishment of data mining experiments.